\documentstyle[12pt]{article}

\def\bq{\begin{quotation}}
\def\eq{\end{quotation}}

\def\fnote#1#2 {\begingroup \def \thefootnote {#1}
\footnote{#2}\addtocounter{footnote}{-1}\endgroup}

\newcommand{\support} {This work is supported in part by U.S. Department of
Energy Contract No.  DE-AC02-76ER13065.}

\newcommand{\myname} {\vspace{0.5in}
                                  \begin{center}Zhu Yang\\
                                  \vspace{0.2in}
                                  Department of Physics and Astronomy\\
                                  University of Rochester\\
                                  Rochester, NY 14627\\
                                  \vspace{0.5in}
                                 Abstract\\
                                  \vspace{0.2in}
                                \end{center}}

\newcommand{\pagenumber}{\pagestyle{plain}\setcounter{page}{1}}

\def\g{\gamma} \def\G{\Gamma}
\def\a{\alpha} 
\def\b{\beta} 
\def\d{\delta} 
\def\e{\epsilon}

 \def\L{\Lambda}
\def\m{\mu} 
\def\n{\nu} 
 
 \def\P{\Pi}
\def\r{\rho} 
\def\s{\sigma} 
\def\t{\tau}

\def\raisenot{\raise .5mm\hbox{/}}
\newcommand{\notpa}{\hbox{{$\partial$}\kern-.54em\hbox{\raisenot}}}
\def\notp{\ \hbox{{$p$}\kern-.43em\hbox{/}}}
\def\notq{\ \hbox{{$q$}\kern-.47em\hbox{/}}}
\def\notk{\ \hbox{{$k$}\kern-.47em\hbox{/}}}
\def\notA{\ \hbox{{$A$}\kern-.47em\hbox{/}}}
\def\nota{\ \hbox{{$a$}\kern-.47em\hbox{/}}}
\def\notb{\ \hbox{{$b$}\kern-.47em\hbox{/}}}

\begin {document}
\baselineskip=24pt

\pagestyle{empty}

\begin{flushright}
UR-1225\\
ER-13065-678
\end{flushright}
\vspace{0.5in}
\begin{center}
{\Large
Effective Superstrings}
\end{center}

\myname

We generalize the method of quantizing effective strings proposed by
Polchinski and Strominger to superstrings. The Ramond-Neveu-Schwarz
string is different from the Green-Schwarz string in non-critical
dimensions. Both are anomaly-free and Poincare invariant.
Some implications of the results are discussed. The formal
analogy with 4D (super)gravity is pointed out.
\newpage
\pagenumber

\section{Introduction}

Recently Polchinski and Strominger have made substantial progress in
quantizing long effective strings such as Nielsen-Olesen vortex and long
QCD flux tubes. The essential point is that the measure in path
integral approach should be built out of the induced metric rather than
the intrinsic metric as used in Polyakov quantization. Such an
assumption is physically appropriate for string solitons that
arise from underlying well-defined dynamics like the Abelian
Higgs model. The results of \cite{pol} are likely to have
impact on quantization of  other effective extended objects.
In this note we consider effective string theories that involve fermionic
structure on the world sheet. For example, a supersymmetric
version of Nielsen-Olesen vortex exists\cite{div}. If we take the case where
there is partially broken
global supersymmetry\cite{hughes}
 with the string soliton, there will
be one world-sheet massless
fermionic degree of freedom accompanying each transverse oscillation
of the string.  Such a string is described by
a Green-Schwarz (GS) string \cite{gs} whose quantization in
sub-critical dimension is unexplored. Another case of interest is
3-dimensional Ising model, which has been shown to be a fermionic
string. In non-critical region, we expect to
consider quantization of an effective  fermionic string, possibly
Ramond-Neveu-Schwarz (RNS) string\cite{dot}.
Our approach is incapable of describing
the critical behavior, or what is the same, continuum
limit, due to the lack of renormalizability.
Finally, we might cite supersymmetric large $N_{c}$ QCD as an example
of effective superstring.
We will keep in mind these examples when we quantize the effective
superstrings.

As pointed out in \cite{pol}, for these strings none of
the standard quantization methods
is correct.
We will follow closely lines of thought and notation in \cite{pol}.
The idea is quite simple: whenever there appears the intrinsic
Liouville field, we replace it by the induced metric and then
check the anomaly cancellation. Field equations are changed and
spectrum modified.
We first work out  the improved superconformal algebra and the spectrum of the
RNS string, since it is more familiar. Next we quantize
GS strings in $D=3,4,6$ and $10$, where they exist.
We develop an operator formulation that incorporates the non-trivial
feature of conformal anomaly of $\theta$ fields.
It turns out that even after GSO projection of the RNS string, the two are
not the same, except when $D=10$.
Finally we point out some  formal analogies of the approach initiated
in \cite{pol}
with other problems in field theory and quantum gravity.

\section{Spinning String}

Locally gauge invariant formulation of a spinning string  to the lowest
order in derivatives is
a 2D conformal supergravity.
It is not known whether it can arise as an effective string from
some underlying microscopic model,
although there is a conjecture
that identifies D=3 spinning string with 3D Ising model.
We quantize this string theory
because of its elegant structure and
for the sake of comparison with the Green-Schwarz string.
To describe a long spinning string, if it can exist at all, one should
use the induced metric and gravitino. They can be found from the
classical equations of motion.

Consider now the RNS string action \cite{deser},
\begin{equation}
S=-{\frac{1}{2\pi a^{2}}}\int d^{2}\s  e \lbrace
h^{\a\b}\partial_{\a} X^{\m} \partial_{\b}X_{\m}-i\bar{\psi}^{\m}\r^{\a}
\nabla_{\a} \psi_{\m} -2\bar{\chi}_{\a}
\r^{\b} \r^{\a} \psi^{\m} \partial_{\b} X_{\m}
-{\frac{1}{2}}\psi_{\m}\psi^{\m}\bar{\chi}_{\a}\r^{\b}\r^{\a}\chi_{\b}\rbrace,
\end{equation}
which is locally supersymmetric with minimal number of derivatives,
thus is the most relevant term of an effective spinning string.
Besides general covariance and local supersymmetry, (1) is also invariant
under conformal transformation
\begin{equation}
\d X^{\m}=0, \, \d \psi^{\m}=-{\frac{1}{2}}\L \psi^{\m},\, \d e_{\a}^{a}
= \L e_{\a}^{a}, \,\d \chi_{\a}={\frac{1}{2}}\chi_{\a}
\end{equation}
and its super partner
\begin{equation}
\d \chi_{\a} = i\r_{\a}\eta, \,\d e^{a}_{\a}=\d \psi^{\m}=\d X^{\m}=0.
\end{equation}
In the spirit of \cite{pol}, the zweibein $e_{\m}^{a}$ and
the gravitino $\chi_{\m}$ are not independent dynamical variables.
Rather they are formed from the matter fields $X_{\m}$ and $\psi_{\m}$.
As in bosonic string, one should vary $e_{\m}^{a}$ and $\chi_{\m}$ and
solve them in terms of $X_{\m}$ and $\psi_{\m}$.
For this purpose we write down the relevant equations of motion,
\begin{equation}
T^{\a}_{a} \equiv {\frac{1}{e}} {\frac{\d I}{\d e^{a}_{\a}}}=0,\,
J_{\a}={\frac{1}{2}}\partial_{\b}X_{\m}\r^{\b}\r^{\a}\psi^{\m}
+{\frac{1}{4}}\bar{\psi}^{\m}\psi_{\m}\r^{\b}\r^{\a}\chi_{\b}=0.
\end{equation}
It is evident from (1) and (4) that we can read off the metric $g_{\m\n}$,
\begin{equation}
g_{\a\b} \sim \partial_{\a}X^{\m}\partial_{\b}X_{\m} + {\frac{1}{2i}}
\bar{\psi}^{\m}\r_{(\a}\nabla_{\b)}\psi_{\m}+ \chi-dependence.
\end{equation}
It is however non-trivial to find $\chi_{\m}$ in general case.
As usual things simplify in superconformal gauge, which means in our case
\begin{equation}
T_{++}=T_{--}=0, \,\, J_{{\frac{1}{2}}+}=J_{-{\frac{1}{2}}-}=0,
\end{equation}
where we have switched to light-cone coordinates and
${\frac{1}{2}}(-{\frac{1}{2}})$ denotes upper (lower) component of
a spinor.
It is easy to see that in this gauge $g_{++}=g_{--}=0$ and
the induced ``Liouville" field is
\begin{equation}
\phi=\ln g_{+-} = \ln [\partial_{+}X^{\m}\partial_{-}X_{\m}+
\psi_{{\frac{1}{2}}}\partial_{-}\psi_{{\frac{1}{2}}}+\psi_{-{\frac{1}{2}}}
\partial_{+}\psi_{-{\frac{1}{2}}}+\chi-dependence].
\end{equation}
In order to find its superpartner $\chi$ we use the following trick.
A local supersymmetry transformation turns $\phi$ into $\chi$:
\begin{equation}
\d X^{\m}=\bar{\e}\psi^{\m}, \d \psi^{\m}=-i\r^{\a}\e(\partial_{\a}X^{\m}
-\bar{\psi}^{\m}\chi_{\a}), \d e^{a}_{\a}=-2i\bar{\e}\r^{\a}\chi_{\a},
\d \chi_{\a}=\nabla_{\a}\e.
\end{equation}
Since
$\phi$ is determined from (7), we can vary (7) to get $\chi$ in terms
of $X_{\m}$ and $\psi_{\m}$. They are
\begin{eqnarray}
\chi_{-{\frac{1}{2}}} &\equiv& \chi_{{\frac{1}{2}}-}=
{\frac{\partial_{-}\psi_{-{\frac{1}{2}}}^{\m}\partial_{+}X_{\m}}
{\partial_{+}X^{\m}\partial_{-}X_{\m}}}+\cdots, \nonumber \\
\chi_{{\frac{1}{2}}} &\equiv& \chi_{-{\frac{1}{2}}+}=
{\frac{\partial_{+}\psi_{+{\frac{1}{2}}}^{\m}\partial_{-}X_{\m}}
{\partial_{+}X^{\m}\partial_{-}X_{\m}}}+\cdots.
\end{eqnarray}
In deriving (9) we have omitted higher orders in fermion fields and
used the $\chi$-dependent terms in (7) to cancel some unwanted
terms linear in $\psi^{\m}$. By the way, a superspace formulation
may facilitate our derivations. We will stick to component form
in this paper.

With these preliminary findings we are in a position to compute anomalies.
It is well-known that the Polyakov determinant\cite{poly,mar}
 in terms of intrinsic
$\phi$ and $\chi$  induces the following action
\begin{equation}
S_{L} = {\frac{D-10}{8\pi}} \int d^{2}\s\left[\partial_{+}\phi\partial_{-}\phi
+\chi_{\frac{1}{2}}\partial_{-}\chi_{\frac{1}{2}}+\chi_{-{\frac{1}{2}}}
\partial_{+}\chi_{-{\frac{1}{2}}}\right].
\end{equation}
Now following Polchinski and Strominger, who in turn used an idea of
DDK\cite{dav}, we simply substitute  (7) and (9) into (10) and get
\begin{eqnarray}
S_{L} &=& {\frac{\b}{\pi}}
  \int d^{2}\s [ {\frac{\partial_{+}^{2}X\cdot\partial_{-}X
\partial_{-}^{2}X\cdot\partial_{+}X}{(\partial_{+}X\cdot\partial_{-}X)^{2}}}
+{\frac{\partial_{+} \psi_{\frac{1}{2}}\cdot\partial_{-}^{2}X}{(\partial_{+}X
\cdot\partial_{-}X)^{2}}} \partial_{+}\psi_{\frac{1}{2}}\cdot\partial_{-}X
\nonumber \\
&+& {\frac{\partial_{-}
\psi_{-{\frac{1}{2}}}\cdot\partial_{+}^{2}X}{(\partial_{+}X
\cdot\partial_{-}X)^{2}}} \partial_{-}\psi_{-{\frac{1}{2}}}\cdot\partial_{+}X].
\end{eqnarray}
As is the case in \cite{pol}, the expectation value of $\partial_{+}X
\cdot \partial_{-}X$ must be non-zero and large compared
to $a^{2}$, i.e., we are dealing with
a long string.
Note in (11) we have introduced a new coupling constant $\beta$ which
will be determined by the anomaly-free condition.
Now our problem is transformed into investigating the dynamics generated
by (1) plus (11).

In order to stabilize our long string, we follow \cite{pol} to
periodically identify space in the $X^{1}$ direction with a large radius
$R$ and consider strings with winding number $1$,
\begin{equation}
X^{\m}(\tau,\s+2\pi)=X^{\m}(\tau,\s)+2\pi R \d^{\m}_{1}.
\end{equation}
The classical background is
\begin{equation}
X_{cl}=e_{+}^{\m}R \s^{+} + e_{-}^{\m}R \s^{-}, \psi_{cl}=0,
\end{equation}
where $e_{+}\cdot e_{+}=e_{-}\cdot e_{-}=0$ and $e_{+}\cdot e_{-}=-1/2$.
We then make an expansion in powers of $R^{-1}$. Write $Y^{\m}=X^{\m}-
X^{\m}_{cl}$, we find
to the next to leading order on $R^{-1}$
\begin{equation}
L=-{\frac{R^{2}}{8\pi a^{2}}}+{\frac{1}{4\pi a^{2}}}\partial_{+}Y\cdot
\partial_{-}Y +
{\frac{1}{2\pi}}
\psi_{\frac{1}{2}}\partial_{-}\psi_{\frac{1}{2}}+
{\frac{1}{2\pi}}\psi_{-{\frac{1}{2}}}
\partial_{+}\psi_{-{\frac{1}{2}}}+ {\frac{\b}{\pi R^{2}}}\partial_{+}^{2}Y
\cdot e_{-} e_{+}\cdot \partial_{-}^{2}Y + O(R^{-3}).
\end{equation}
Note that the fermionic terms in (11) do not contribute in this order.
Although not obvious, (14) still possesses superconformal invariance
order to order in $R^{-1}$.
For convenience we will first compute the modified (super) stress tensor, then
use operator product expansion (OPE) to obtain new transformation laws for the
matter fields.
The superstress tensor can be simply read off from
the expression of the intrinsic Liouville contribution.
According to \cite{dis}, it is
\begin{eqnarray}
T_{++} &=&{\frac{1}{2}}\partial_{+}X\cdot\partial_{+}X+
{\frac{1}{2}}\psi_{\frac{1}{2}}
\partial_{+}\psi_{\frac{1}{2}}+
{\frac{\b}{2}} (\partial_{+}\phi\partial_{+}\phi+
\partial_{+}^{2}\phi+\chi_{\frac{1}{2}}
\partial_{+}\chi_{\frac{1}{2}}),\nonumber \\
G_{\frac{3}{2}} &=& {\frac{1}{2}}\partial_{+}X\cdot \psi_{\frac{1}{2}}+
{\frac{\b}{2}}(\partial_{+}\chi+\partial_{+}\phi \chi_{\frac{1}{2}}).
\end{eqnarray}
Now we substitute (7) and (9) into the above equation and get
\begin{eqnarray}
T_{++} &=& -{\frac{R}{a^{2}}}e_{+}\cdot \partial_{+}Y
-{\frac{1}{2}}\partial_{+}Y\cdot\partial_{+}Y+{\frac{1}{2}}\psi_{\frac{1}{2}}
\partial_{+}\psi_{\frac{1}{2}}-{\frac{2\b}{R}}e_{-}\cdot\partial_{+}^{3}Y,
\nonumber \\
G_{\frac{3}{2}} &=& {\frac{1}{2}}Re_{+}\cdot \psi_{\frac{1}{2}}+
 {\frac{1}{2}}\partial_{+}Y\cdot \psi_{\frac{1}{2}}+
{\frac{2\b}{R}} e_{-}\cdot \partial_{+}^{2}\psi_{\frac{1}{2}}+(R^{-1}).
\end{eqnarray}
Note that to order $R^{-1}$ only the improvement term contributes.
This procedure of getting $T_{++}$ and $G_{\frac{3}{2}}$
 resembles the so-called 1.5 order formalism of supergravity
\cite{van}.
It is then straightforward to evaluate the OPEs of stress tensor and
its superpartner. We obtain,  to order $R^{0}$,
\begin{eqnarray}
T_{++}(\s^{+}) T_{++}(0) &=& {\frac{3}{4}} {\frac{D-8\b}{(\s^{+})^{4}}}
+{\frac{2}{(\s^{+})^{2}}} T_{++}(0)+{\frac{1}{\s^{+}}}\partial_{+}T_{++}(0)
+O(R^{-1}),\nonumber \\
T_{++}(\s^{+}) G_{\frac{3}{2}}(0) &=& {\frac{3}{2}}{\frac{1}{(\s^{+})^{2}}}
G_{\frac{3}{2}}(0)+{\frac{1}{\s^{+}}}\partial_{+}G_{\frac{3}{2}}(0)+ O(R^{-1}),
\nonumber \\
G_{\frac{3}{2}}(\s^{+}) G_{\frac{3}{2}}(0) &=& {\frac{D-8\b}{4(\s^{+})^{3}}}
+{\frac{1}{2\s^{+}}}T_{++}(0)+O(R^{-1}).
\end{eqnarray}
If $\beta=(D-10)/8$, the  anomaly in (17) cancels the ghost contribution.
As in bosonic case, $\beta$ is not renormalized.
As usual, the superconformal transformation law of $X^{\m}$ and $\psi^{\m}$
is found from the OPE between them and T and G.
Since
\begin{eqnarray}
T_{++}(\s^{+})Y^{\m}(0) &=& {\frac{1}{\s}}(Re^{\m}_{+}+\partial_{+}Y^{\m})
+{\frac{1}{(\s^{+})^{3}}} {\frac{2\b e_{-}^{\m}}{R}}+\cdots,\,
T_{++}(\s^{+})\psi_{\frac{1}{2}}^{\m}(0)={\frac{1}{2\s^{+}}}
\psi_{\frac{1}{2}}^{\m}+\cdots,\nonumber \\
G_{\frac{3}{2}}(\s^{+}) Y^{\m}(0) &=&
{\frac{1}{2\s^{+}}}\psi_{\frac{1}{2}}^{\m}
+\cdots,\, G_{\frac{3}{2}}(\s^{+}) \psi_{\frac{1}{2}}^{\m}(0)=
{\frac{1}{2\s^{+}}}(Re_{+}^{\m}+\partial_{+}Y^{\m})+
{\frac{1}{(\s^{+})^{3}}}{\frac{4\b e_{-}^{\m}}{R}},
\end{eqnarray}
we have, under left-moving (super) conformal transformation $\e^{+}(\s^{+})$,
and $\eta_{-\frac{1}{2}}(\s^{+})$,
\begin{eqnarray}
\d Y^{\m} &=& \e^{+}(Re_{+}^{\m}+
\partial_{+}Y^{\m})-{\frac{\b}{R}}\partial_{+}^{2}
\e^{+} e_{-}^{\m}+{\frac{1}{2}}\eta_{-\frac{1}{2}}\psi_{\frac{1}{2}},\,
\nonumber \\
\d \psi_{\frac{1}{2}}^{\m} &=& {\frac{1}{2}}\partial_{+}\e^{+}
\psi_{\frac{1}{2}}^{\m}+\e^{+}\partial_{+}\psi_{\frac{1}{2}}^{\m}
+\eta_{-\frac{1}{2}}(Re_{+}^{\m}+\partial_{+}Y^{\m})+
{\frac{2\b}{R}}\partial_{+}^{2}\eta_{-\frac{1}{2}} e_{-}^{\m}.
\end{eqnarray}

We now examine the spectrum.
The bosonic oscillations $Y^{\m}$ have harmonic expansion with periodic
boundary conditions,
\begin{equation}
\partial_{+}Y=a \sum^{+\infty}_{n=-\infty} \a^{\m}_{n}e^{-in\s^{+}}.
\end{equation}
For $\psi^{\m}$, there are two different expansions as usual:
one with periodic boundary condition (R sector)
\begin{equation}
\psi_{+}^{\m}={\frac{a}{\sqrt{2}}} \sum_{n \in Z} d_{n}^{\m} e^{-in\s^{+}},
\end{equation}
and one with antiperiodic condition (NS sector)
\begin{equation}
\psi_{+}^{\m}={\frac{a}{\sqrt{2}}} \sum_{n \in Z+{\frac{1}{2}}}
d_{n}^{\m} e^{-in\s^{+}},
\end{equation}
Insert the above into the mode expansion of $T$ and $G$ we have the
superconformal generators
\begin{eqnarray}
L_{n} &=& {\frac{R}{a}}e_{+}\cdot\a_{n}+{\frac{1}{2}}\sum_{m\in Z}
:\a_{n-m}\a_{m}: -
{\frac{\b}{R}}a n^{2}e_{-}\cdot \a_{n}
+ L_{m}^{f}, \nonumber \\
G_{r} &=& R e_{+}\cdot b_{r}+\sum_{n=-\infty}^{\infty}\a_{-n}\cdot b_{r+n}
+{\frac{8\b r^{2}}{R}}e_{-}\cdot b_{r}, \nonumber \\
F_{m} &=& R e_{+}\cdot d_{m} + \sum_{n=-\infty}^{\infty} \a_{-n}\cdot d_{m+n}
+{\frac{8\b m^{2}}{R}} e_{-}\cdot d_{m},
\end{eqnarray}
where $f=b,d$ denotes NS, R sector contribution to $L_{n}$, respectively:
\begin{eqnarray}
L_{n}^{b} &=& {\frac{1}{2}}\sum_{r=-\infty}^{\infty}(r+{\frac{1}{2}}n)
:b_{-r}\cdot b_{n+r}: -{\frac{\b}{2}}\d_{n,0}, \nonumber \\
L_{n}^{d} &=& {\frac{1}{2}}\sum_{m=-\infty}^{\infty}(m+{\frac{1}{2}}n)
:b_{-n}\cdot b_{m+n}:.
\end{eqnarray}
They satisfy the superconformal algebra with central charge 10,
\begin{eqnarray}
[L_{m},L_{n}] &=& (m-n)L_{m+n}+A(m)\d_{m,0}, \nonumber \\
\left[L_{m},G_{r}\right] &=& ({\frac{1}{2}}m-r)G_{m+r},\nonumber \\
\lbrace G_{r}, G_{s} \rbrace &=& 2L_{r+s}+ B(r)
\d_{r+s,0},
\end{eqnarray}
where
\begin{equation}
A(m)={\frac{5}{4}}m(m^{2}-1),\,\, B(r)=5(r^{2}-{\frac{1}{4}}).
\end{equation}
For $R$ sector, we replace $G_{r}$ by $F_{n}$ and
similar structure emerges with
\begin{equation}
A(m)={\frac{5}{4}}m^{3},\,\, B(n)=5n^{2}.
\end{equation}
Now we discuss the ground state. It is well-known that the tachyonic
ground state of fundamental RNS string in critical dimension
$D=10$  and half of the excited states are inconsistent with modular invariance
and thus must be projected out by GSO projection\cite{gso}.
As a bonus space-time
supersymmetry is realized. Is it necessary to do GSO projection for
effective spinning string?
In the case of Nielsen-Olesen string or its supersymmetric generalization,
the string is stable. Interaction like breaking and joining are not allowed.
In this case, presumably GSO projection is not necessary.
If string interaction exists, one must be prepared to do GSO projection
to achieve consistency.
In any case, before the  GSO projection, the on-shell condition for the NS
sector is
\begin{equation}
G_{r} |\phi\rangle = 0 \, (r>0), L_{n}|\phi\rangle= 0 \, (n>0),
(L_{0}-{\frac{1}{2}})|\phi\rangle=0,
\end{equation}
while for the R sector it is
\begin{equation}
F_{n}|\phi\rangle =L_{n}|\phi\rangle=0 \, (n\ge 0).
\end{equation}
They are the same as for critical strings because they are consequences
of, say, BRST invariance \cite{gsw}.
For the NS ground state $|k,k\rangle$, the total momentum is
\begin{equation}
p^{\m}={\frac{R}{2a^{2}}}(e^{\m}_{+}+e^{\m}_{-})+{\frac{1}{2}}(\a_{0}^{\m}
+\tilde{\a}_{0}^{\m}).
\end{equation}
{}From the on-shell condition (25) we have $k^{1}=0$.
So the mass is
\begin{equation}
m= {\frac{R}{2a^{2}}}- {\frac{D-2}{8R}}.
\end{equation}
It is a scalar.
For R sector, the ground state energy is simply $R$.
The reason is that to the leading order, there is no quantum correction to
$F_{0}$.

It is instructive to work out the first excited states in NS sector and
observe whether they pair with R ground state to form a supermultiplet.
Let's look at the following state
\begin{equation}
|\phi\rangle = \e \cdot b_{-\frac{1}{2}} \tilde{\e}
\cdot \tilde{b}_{-\frac{1}{2}} |k,\tilde{k};0\rangle,
\end{equation}
where $\e^{\m}$ and $\tilde{\e}^{\m}$ are the polarization vectors.
The $G_{\frac{1}{2}}|\phi\rangle=0$ implies that
$\e \cdot v=0$,  where $v^{\m}= R e_{+}^{\m}+k^{\m}+\b e_{-}^{\m}/4R$.
{}From $(L_{0}-1/2)|\phi\rangle=0$ we have,
$v^{\m}v_{\m}=0$. The two conditions imply that there are $D-2$ physical
components for each $\e$ and $\tilde{\e}$.
The rest mass is $\sqrt{-p^{2}}=R/2a^{2}-(D-10)/8R$, which
is different from the R ground state.
This suggests that there is no space-time supersymmetry in transverse
directions even
if we take the GSO
projection, except for $D=10$.
In the next section we will study Green-Schwarz string, where
transverse space supersymmetry survives quantization.
It is somewhat surprising that the two formulations of fermionic
string differ in non-critical dimensions.

\section{Green-Schwarz superstring}

The Green-Schwarz string describes space-time supersymmetric string.
In covariant formulation, the action reads\cite{gs}
\begin{eqnarray}
S &=& -{\frac{1}{2}}\int d^{2}\s \lbrace
\sqrt{g} g^{\a\b}\P_{\a}\cdot\P_{\b}+
+2i\e^{\a\b} \partial_{\a}X^{\m}(\bar{\theta}^{1}\G_{\m}\theta^{1}
-\bar{\theta}^{2}\G_{\m}\theta^{2})\, \nonumber \\
&+&\e^{\a\b}\bar{\theta}^{1}\G^{\m}
\partial_{\a}\theta^{1}\bar{\theta}^{2}\G_{\m}\partial_{\b}\theta^{2}\rbrace,
\nonumber \\
\Pi_{\a}^{\m} &=& \partial_{\a}X^{\m}-i\bar{\theta}^{A}\G^{\m}
\partial_{\a}\theta^{A}.
\end{eqnarray}
The nonlinearity and peculiar constraints have made an straightforward
covariant quantization impossible. In fact GS string is most thoroughly
studied only in light-cone gauge operator formulation, where the modification
of the theory in non-critical dimension is not known. A world-sheet covariant
path integral method has been developed and anomaly cancellation
mechanism is understood\cite{carlip,kal}.
Very little is known, however, about the non-critical
GS string.
In view of the weak coupling supersymmetric Abelian Higgs model, however,
a consistent quantization of a long string definitely exists.
Moreover, the  powerful  method of conformal field theory has not
been applied to  the GS string. Of course the path integral
approach and CFT must be equivalent, the latter is however easier to handle.
In what follows, we will show how the conformal anomaly
arises and differs from familiar first order systems in  the fermion sector
in the language of OPE, and how to construct an anomaly-free non-critical
theory. As we will see, the interaction between bosons and fermions
 is essential.

For a long string, a non-covariant gauge fixing in \cite{hughes,carlip,kal}
is natural and
satisfactory because the string itself breaks Lorentz invariance
spontaneously.
The gauge fixing is
\begin{equation}
\g^{+}\theta^{A}=0, \g_{\a\b}=e^{\phi}\eta_{\a\b}.
\end{equation}
The first one in (34) fixes the Siegel $\kappa$-symmetry\cite{sie}.
In this gauge the classical action (33) simplifies to
\begin{equation}
S=\int d^{2}{\s}\lbrace {\frac{1}{2}}\partial_{+}X^{\m} \partial_{-}X_{\m}
+ \bar{\theta}^{1}\g^{-}\partial_{+}X^{+}\partial_{-}\theta^{1}
+ \bar{\theta}^{2}\g^{-}\partial_{-}X^{+}\partial_{+}\theta^{2}\rbrace.
\end{equation}
The stress tensor is
\begin{equation}
T_{++}={\frac{1}{2}}\Pi_{+}\cdot\P_{+}.
\end{equation}
As it stands, $T_{++}$ is not well-defined quantum mechanically, due
to short distance singularity. Conventionally one normal orders
the product of two operators. In case of interacting fields
or  in the presence of external fields, the
normal ordering usually produces finite terms called anomaly.
This happens to $T_{++}$ too. We normal order
one of the terms in (36), $\partial_{+}X^{\m}
 \bar{\theta}\g^{\m}\partial_{+}\theta$, since it is the only relevant
one contributing to conformal anomaly.
The $\theta^{1}$ propagator reads
\begin{equation}
\bar{\theta}^{1}(\s^{+})\, \theta^{1}(0) = {\frac{1}{2
\g^{-}\sqrt{\partial_{+}X^{+}(\s^{+})} \s^{+} \sqrt{\partial_{+}X^{+}(0)}}},
\end{equation}
where $\g^{-}$ is invertible in the subspace of $\g^{+}\theta^{1}=0$.
The particular form of (37) is chosen to be consistent with
general  covariance.
We must replace $\partial_{+}X^{\m}
 \bar{\theta}^{1}\g^{\m}\partial_{+}\theta^{1}$ by
\begin{equation}
\partial_{+}X^{\m}
: \bar{\theta}\g^{\m}\partial_{+}\theta: - {\frac{N_{f}}{8}}
{\frac{\partial_{+}X^{+}\partial_{+}^{3}X^{+}}
{(\partial_{+}X^{+})^{2}}}+ O(R^{-2}),
\end{equation}
where where as before $\partial_{+}X^{+} \sim R$
and  $N_{f}$ is number of propagating
$\theta$ components. Here the
normal product simply means that there is no contraction between
fields inside it.
The significance of the anomalous term in (38)
is that it will produce extra conformal anomaly
for $\theta$, because  we are  effectively treating $\theta$ as if it had
conformal dimension 1/2, but actually it has dimension 0.
Another noteworthy feature of it is its resemblance
to the quantum contribution to the stress tensor (16).
The normal ordering of other terms in (36) makes no contribution to the leading
order.
Now we write $T_{++}$ we have got so far, again in the background
(12) and (13),
\begin{equation}
T_{++}= - {\frac{R}{a^{2}}} e_{+}\cdot \partial_{+}Y-
{\frac{1}{2}}\partial_{+}Y\cdot\partial_{+}Y+ e_{+}^{\m} :\bar{\theta}^{1}
\g^{\m}\partial_{+}\theta^{1}:- {\frac{N_{f}}{8}}
{\frac{1}{Re_{+}\cdot u}} \partial_{+}^{3}Y^{+}
+O(R^{-2}),
\end{equation}
where $u^{\m}$ is the unit vector in $X^{+}$ direction.
If we compute the OPE of $T_{++}$'s we find that
the anomaly is
\begin{equation}
c=D+2N_{f},
\end{equation}
 Here, $D$ is of course the $X^{\m}$ contribution.
$N_{f}/2$  is from direct $\theta$ contraction, and the remaining
$3N_{f}/2$ is due to the contraction  between the first
 and the last term of (39).
For $D=10,6,4,3$, $N_{f}=8,4,2,1$, respectively.
Now follow the previous treatment\cite{pol}, we add to $T_{++}$ a ``Liouville"
contribution $\b e_{-}\cdot \partial_{+}^{3}/R$. The complete
quantum  $T_{++}$ then reads, up to $O(R^{-2})$,
\begin{equation}
T_{++}= - {\frac{R}{a^{2}}} e_{+}\cdot \partial_{+}Y-
{\frac{1}{2}}\partial_{+}Y\cdot\partial_{+}Y+ e_{+}^{\m} :\theta^{1}
\g^{\m}\partial_{+}\theta^{1}:+ c {\frac{1}{Re_{+}\cdot
u}}\partial_{+}^{3}Y^{+}
+{\frac{\b}{R}} e_{-}\cdot \partial_{+}^{3}Y+
O(R^{-2}),
\end{equation}
where $\b=(26-c)/12$. As before, one can check that the anomaly
adds up to 26. Thus we have constructed an anomaly-free effective
Green-Schwarz string. In particular, $\b=0$ for $D=10$.
The path integral evaluation of conformal anomaly in critical GS string
\cite{carlip,kal}
is recovered.

Let's examine the spectrum. It is straightforward to make mode expansion
of $T_{++}$ in terms of $L_{n}$  and obtain the Virasoro algebra with
central charge 26.
As usual, a physical state $|\phi\rangle$
 is defined as
\begin{equation}
L_{n} |\phi\rangle=0 (n>0),\,\,\, (L_{0}-1)|\phi\rangle=0.
\end{equation}
An important issue is to count the contribution of Casimir
effect to $L_{0}$.
We have $D/24$ from $D$ bosons, and $-2$ from  the reparametrization ghosts.
For a closed string, the only boundary condition for $\theta$
is periodicity,
\begin{equation}
\theta^{A}(\s,\t) =\theta^{A}(\s+2\pi,\t).
\end{equation}
So
\begin{equation}
\theta^{1}=\sum_{n\in Z} \theta^{1}_{n}e^{-2in\s^{+}}.
\end{equation}
Since a periodic fermion has Casimir energy $-1/24$, we have
$-N_{f}/24$ from $\theta$. For $D=3,4,6,10$, the Casimir energy all cancels.
Thus we write
\begin{equation}
L_{0} ={\frac{R}{a}}\e_{+}\cdot\a_{0}+{\frac{1}{2}}\sum_{m\in Z}
:\a_{-m}\cdot
\a_{m}:+ e_{+}^{+}\sum_{m\in Z}:n\bar{\theta}_{-n}\g^{-}\theta_{n}:
+O(R^{-2}).
\end{equation}
The ground state thus has mass $R$, as the R sector of RNS string.
The ground state forms  representations of the $\theta$ zero
modes, which is expected to generate supersymmetry in the transverse
direction. We leave the details for further study.
To conclude this section, we have achieved a world-sheet covariant quantization
of non-critical Green-Schwarz string. The ground state energy
is not modified by quantum effects, at least to the leading nontrivial
order we are studying. This suggests  that the transverse
supersymmetry persists
after quantization.

\section{Discussion}
A. As we have seen the RNS and GS strings differ from each other when $D\ne
10$.
This can be understood in terms of degrees of freedom counting.
Th ground state energy (31) of the NS sector is a simple consequence
of transversality of physical excitations. Since the Regge slope
does not depend on $D$, the quantum correction can not vanish for every
$D$. On the other hand, in GS string, the number of
bosonic physical degrees of freedom $(D-2)$ equals exactly that of fermions,
resulting the cancellation in (45). If we use light-cone
gauge quantization, the equivalence proof of the RNS and GS strings in
\cite{witten}
must break down in $D\ne 10$, possibly due to the lack of triality of
 $Spin(D-2)$.

B. The continuum limits of the two strings, it they exist, must be different.
This raises an interesting question concerning the continuum limit of
$3D$ Ising model. It is known that the $3D$ Ising model can be represented as
a Fermionic string theory. For aesthetic reason and simplicity, one conjectures
that the string should have maximal symmetry. This leads to RNS and GS strings.
The question is then, which one describes the Ising model?

C. Kutasov and Seiberg\cite{kut}
 have constructed tachyon-free RNS strings with the
intrinsic Liouville field. Their string theory has the space supersymmetry
which corresponds roughly to the supersymmetry in the transverse direction.
It would be interesting to redo their calculation  \`{a} la Polchinski and
Strominger and see whether it is equivalent to the non-critical GS string.
The main difference from our construction of the RNS string is their
insistence on global $N=2$ world sheet supersymmetry.

D. As pointed out by Hughes and Polchinski, there can be a supersymmetric
vortex with the right-left imbalance of the world sheet fermion content.
See also \cite{harvey}.
Such a string is heterotic\cite{gross}.
 The Lorentz anomaly on the world sheet should
be cancelled by some additional fermions, as in the critical case.
The conformal gauge approach to non-critical heterotic string
has been worked out in \cite{yang}. There should be no problem in
applying it to an effective heterotic string.

E. The fact that the two formulations of superstrings are different in
general raises a similar question concerning super p-branes\cite{liu}.
In fact, a supersymmetric generalization \`{a} la RNS must include
the Einstein term\cite{howe}. It is not clear whether it is equivalent
to the GS-like supermembranes. Semiclassical quantization in light-cone
gauge of supermembranes is discussed in \cite{duff}.

F. Polchinski and Strominger have made a nice analog of the effective
string theory to low energy  pion dynamics. Here we make another
analog, this time  with $4D$ Einstein gravity. The general relativity
is based on purely geometrical concept, is
non-renormalizable, and makes sense only when we expand around a
non-singular background. These are the features of effective string
theories too.
In both cases one can add higher derivative terms to
make the theories renormalizable, but at the price of giving up (perturbative)
unitarity (for classical statistical mechanics of a membrane, this
is not a problem).  Moreover, both theories allow supersymmetric
generalization.
It is perhaps then adequate to take the effective string theory as a
laboratory for quantum gravity.
For example, it is possible that the continuum limit of an effective
string theory involves an understanding of expansion around
the unbroken  vacuum $\partial_{\a}X_{cl}=0$.
This in turn might teach us something about the
unbroken phase\cite{witten2}  of general relativity and  fundamental string
theory.

In conclusion, we have quantized the effective superstrings.
The RNS and GS strings are different in non-critical dimensions. We
have  also pointed
out some formal  analogies with 4D (super)gravity.
\vspace{0.25in}
\begin{flushleft}
{\bf Acknowledgement}\\
I would like to thank Joe Polchinski for discussions. \support
\end{flushleft}


\begin{thebibliography}{999}
\bibitem{pol}\, J. Polchinski and A. Strominger, Texas preprint UTTG-17-91.
\bibitem{div}\, P. Di Vecchia and S. Ferrara, Nucl. Phys. {\bf B130}
(1977) 93.
\bibitem{hughes}\, J. Hughes and J. Polchinski, Nucl. Phys. {\bf B278}
(1986) 147.
\bibitem{gs}\, M.B. Green and J.H. Schwarz, Phys. Lett. {\bf 136B} (1984) 367.
\bibitem{dot}\, See, e.g., V.S. Dotsenko, Nucl. Phys. {\bf B285} (1987) 45.
\bibitem{deser}\, S. Deser and B. Zumino, Phys. Lett. {\bf 65B} (1976) 369;\\
L. Brink, P. Di Vecchia and P. Howe, Phys. Lett. {\bf 65B} (1976) 471.
\bibitem{poly}\, A.M. Polyakov, Phys. Lett. {\bf 103B} (1981) 211.
\bibitem{mar}\, E. Martinec, Phys. Rev. {\bf D28} (1983) 2604.
\bibitem{dav}\, F. David, Mod. Phys. Lett. {\bf A3} (1988) 1651;\\
J. Distler and H. Kawai, Nucl. Phys. {\bf B321} (1989) 509.
\bibitem{dis}\, J. Distler, Z. Hlousek and H. Kawai, Int. J. Mod. Phys.
{\bf A5} (1990) 391.
\bibitem{van}\, P. van Nieuwenhuizen, Phys. Rep. {\bf 68} (1981) 189.
\bibitem{gso}\, F. Gliozzi, J. Scherk and D. Olive, Nucl. Phys. {\bf B122}
(1977) 253.
\bibitem{gsw}\, See, e.g., M.B. Green, J.H. Schwarz and E. Witten,
Superstring Theory, Cambridge (1987).
\bibitem{carlip}\, S. Carlip, Nucl. Phys. {\bf B195} (1987) 365.
\bibitem{kal}\, R.E. Kallosh and A.Y. Morozov, Int. J. Mod. Phys. {\bf A3}
(1988) 1943.
\bibitem{sie}\, W. Siegel, Phys. Lett. {\bf 128B} (1983) 397.
\bibitem{witten}\, E. Witten, in Fourth Workshop on Grand Unification,
ed. P. Langacker et al., Birkhauser (1983).
\bibitem{kut}\, D. Kutasov and N. Seiberg, Phys. Lett. {\bf 251B} (1990) 67.
\bibitem{harvey}\, J.A. Harvey, Cargese Lecture (1987).
\bibitem{gross}\, D.J. Gross, J.A. Harvey, E. Martinec and R. Rohm, Phys. Rev.
Lett. {\bf 54} 502.
\bibitem{yang}\, Z. Yang, unpublished (1989).
\bibitem{liu}\, J. Hughes, J. Liu and J. Polchinski, Phys. Lett. {\bf 180B}
(1986) 370.
\bibitem{howe}\, P.S. Howe and R.W. Tucker, J. Phys. {\bf A10} (1977) L155.
\bibitem{duff}\, M.J. Duff, T. Inami, C.N. Pope, E. Sezgin and K.S. Stelle,
Nucl. Phys. {\bf B297} (1988) 515.
\bibitem{witten2}\, E. Witten, Phil. Trans. R. Soc. Lond. {\bf A329} (1989)
349.

\end{thebibliography}
\end{document}